\documentclass[final]{IEEEtran}
\pdfoutput=1
\usepackage{amsthm,amssymb,graphicx,multirow,amsmath,color,amsfonts}
\usepackage[update,prepend]{epstopdf}
\usepackage[noadjust]{cite}
\usepackage[latin1]{inputenc}
\usepackage{tikz}
\usetikzlibrary{arrows,calc}		
\usepackage{bbm} 
\usepackage{pdfpages}
\usepackage{tabulary}
\usepackage{multirow}
\usepackage{comment}
\usepackage{mathtools}
\usepackage{bigints}
\usepackage[inline]{enumitem}
\usepackage{adjustbox}

\def\nb0{{\mathbf{0}}}
\def\nb1{{\mathbf{1}}}

\def\calC{{\mathcal{C}}}

\def\calL{{\mathcal{L}}}

\def\calR{{\mathcal{R}}}

\def\nbbE{{\mathbb{E}}}

\def\nbbR{{\mathbb{R}}}

\def\nrmd{{\rm d}}

\def\nrmx{{\rm x}}
\def\nrmy{{\rm y}}

\newtheorem{lemma}{Lemma}
\newtheorem{thm}{Theorem}

\newtheorem{assumption}{Assumption}
	
\def\P{\mathbb{P}}

\def\sir{\mathtt{SIR}}

\allowdisplaybreaks 

\begin{document}
\bstctlcite{IEEEexample:BSTcontrol}
\graphicspath{{./}}
\title{
On the Load Distribution of Vehicular Users Modeled by a Poisson Line Cox Process
}
\author{
Vishnu Vardhan Chetlur and Harpreet S. Dhillon
\thanks{The authors are with Wireless@VT, Department of ECE, Virginia Tech, Blacksburg, VA (email: \{vishnucr, hdhillon\}@vt.edu). The support of the US NSF (Grant IIS-1633363) is gratefully acknowledged.\hfill
Manuscript last updated: \today.
}
}
\maketitle

\begin{abstract}
	In this letter, we characterize the load on the cellular macro base stations (MBSs) due to vehicular users modeled by a Poisson line Cox process (PLCP). Modeling the locations of MBSs by a homogeneous 2D Poisson point process (PPP), we first characterize the total chord length distribution of the lines of the Poisson line process (PLP) intersecting the typical Poisson Voronoi (PV) cell. Using this result, we derive the exact probability mass function (PMF) of the load on the typical MBS. Considering the computational complexity of this expression, we propose an easy-to-use approximation for the PMF that is also remarkably accurate. Building on this result, we also compute the PMF of the load on the tagged MBS that serves the typical vehicular user. This result enables the characterization of the rate coverage of the typical receiver in the network, which is also included as a useful case study.
\end{abstract}

\section{Introduction} \label{sec:intro}


The third generation partnership project (3GPP) has extended the support for vehicular communications as a part of the fifth generation new radio (5G-NR) standard \cite{3gpp_rel16}. The wide range of use cases supported by vehicular communications includes the exchange of short safety messages with high reliability and also video streaming services with low latency and high data rates. In order to design efficient networks that meet these stringent requirements, it is imperative to understand the additional burden on the network infrastructure due to vehicular users. The distribution of the number of users served by a cellular MBS, which is referred to as the load on the MBS, is a key metric that represents the demand of network resources. 
Inspired by the randomness in the spatial distribution of vehicular users, we employ tools from stochastic geometry to characterize the load on the MBSs due to vehicular users. The prior art in this area is discussed next.

{\em Prior art.} While the distribution of load on MBSs due to users modeled by PPPs has been investigated in the literature \cite{Dhi_letter}, the results cannot be directly applied to vehicular networks because of their unique spatial geometry in which the locations of nodes are
coupled with underlying roadways. This shortcoming of the PPP model was recognized long ago in \cite{bacc_plp}, which proposed a doubly stochastic model to capture this spatial coupling by modeling the layout of roads by a PLP and the locations of vehicles on each of these roads by independent 1D PPPs. This spatial model, referred to as the PLCP, has recently been employed in several works on vehicular communication networks \cite{vishnuJ2, vishnuL1, vishnuJ4,  baccchoi}. The reliability of the typical link in a vehicular ad hoc network (VANET) has been studied in \cite{vishnuL1}. The signal-to-interference ratio (SIR)-based coverage analysis of the typical receiver in a vehicular network has been presented in \cite{vishnuJ2, baccchoi}. In \cite{vishnuJ4}, the coverage analysis of a vehicular network in the presence of shadowing has been studied by leveraging the asymptotic characteristics of the PLCP. In the same work, the authors have also provided the first moment of the load on the MBSs and computed the rate coverage of the typical receiver in a multi-tier vehicular network. However, the characterization of the PMF of the load on the MBSs due to vehicular users modeled by a PLCP remains a key open problem and is the main focus of this letter. More details of our contribution are provided next.

{\em Contributions.} In this letter, we consider a single-tier vehicular network in which the locations of vehicular users are modeled by PLCP and the locations of MBSs serving these users are modeled by a homogeneous 2D PPP. For this setup, we first compute the Laplace transform of the distribution of the total chord length in the typical Poisson Voronoi (PV) cell. We then derive the exact PMF of the load on the typical MBS in terms of the derivative of the Laplace transform of the chord length distribution. Since the evaluation of the exact expression is not computationally efficient, we also propose a simple yet accurate approximation of the PMF of the load on the typical MBS. Using a similar approach, we also derive the PMF of the load on the MBS that serves the typical vehicular user. We also discuss the application of this result in the computation of the rate coverage of the typical receiver in the network. 

\section{System Model}\label{sec:sysmod}

\begin{figure}
	\centering
	\includegraphics[width=0.36\textwidth]{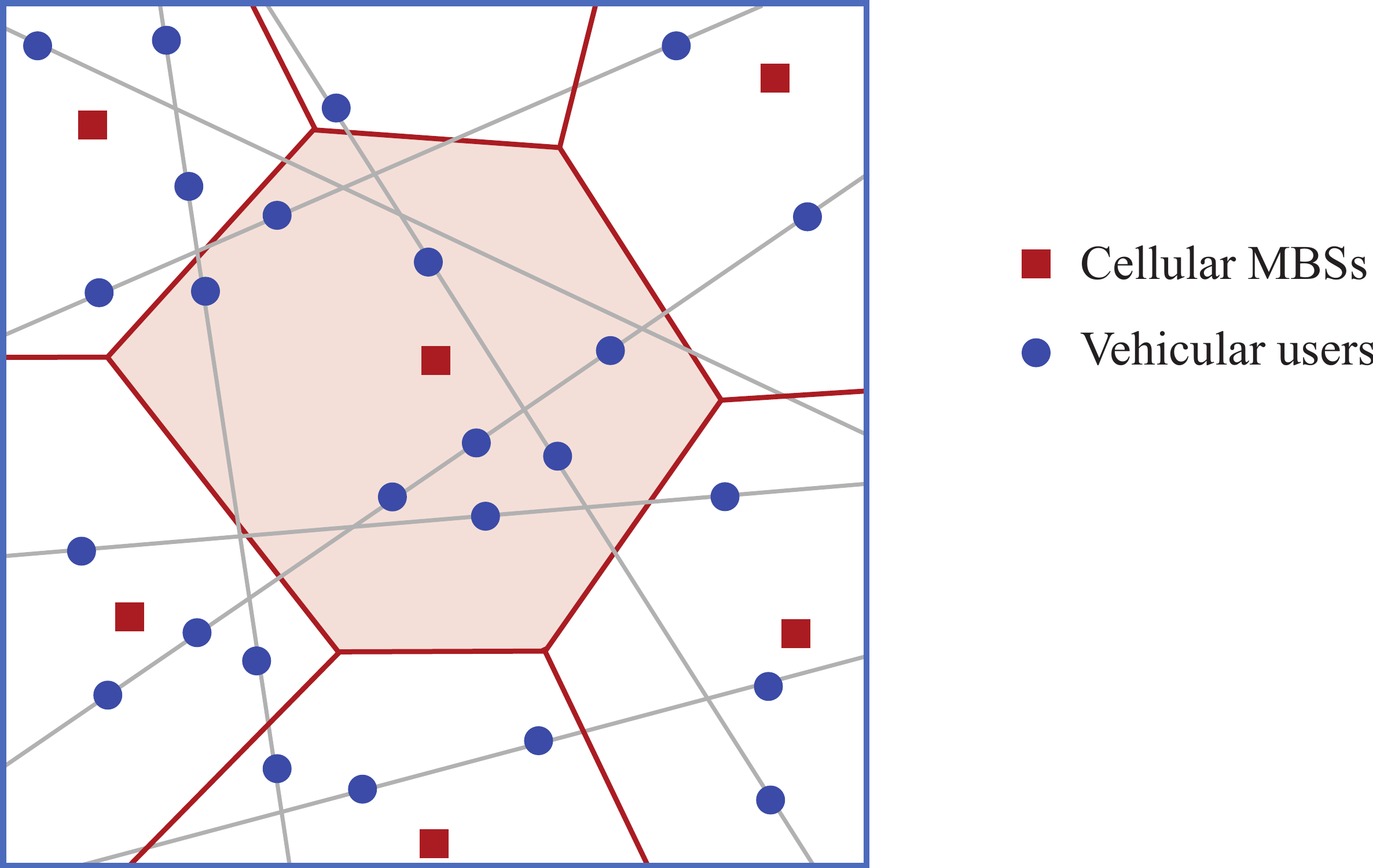}
	\caption{{Illustration of the system model.}}
	\label{fig:sysmod}
\end{figure}

We consider a vehicular network in which the locations of vehicular users are modeled by a motion-invariant PLCP, as illustrated in Fig. \ref{fig:sysmod}. First, we model the physical layout of the roads by a motion-invariant PLP $\Phi_l$ with line density $\mu_l$. We denote the corresponding point process of $\Phi_l$ in the representation space $\calC : \nbbR^+ \times [0, 2\pi)$ by $\Phi_\calC$ which is a homogeneous 2D PPP with density $ \lambda_l = \frac{\mu_l}{\pi} $ \cite{stoyan, vishnuJ2}. Further, we model the locations of vehicular users on each line $L_i \in \Phi_l$ by a homogeneous 1D PPP $\Psi_{L_i}$ with density $\lambda_v$, thereby forming a motion-invariant PLCP $\Phi_v$. We assume that the vehicular users are served by cellular MBSs, whose locations are modeled by a homogeneous 2D PPP $\Phi_b$ with density $\lambda_b$. 

We assume that all the MBSs have the same transmit power. Further, we assume a maximum average received power based association policy and as a result, the association regions of the MBSs are the cells of the Poisson Voronoi tessellation (PVT) formed by $\Phi_b$. For completeness, the PV cell for $\nrmx \in \Phi_b$ is defined as
\begin{align}
V_\nrmx = \{ \nrmy \in \nbbR^2 : \| \nrmy - \nrmx \| \leq \|\nrmy - \nrmx'\|, \forall \nrmx' \in \Phi_b  \}.
\end{align}
The cell $V_o$ formed by the \textit{typical MBS} at the origin is termed the \textit{typical cell}.
Our goal is to characterize the number of vehicular users located within the association region of the MBS. Specifically, we will derive the PMF of the load on the typical MBS and the serving MBS of the typical vehicular user, which will be referred to as the \textit{tagged MBS}. 

\section{Load on the typical MBS}
{In this section, we will derive the PMF of the load on the	typical MBS due to vehicular users, which is one of the key inputs to the network dimensioning process and is also useful in the characterization of volume of traffic at the MBS.}


\subsection{Chord length distribution}

As the locations of vehicular users are confined to roadways, the number of users within the typical cell directly depends on the length of the lines inside the typical cell. So, in order to derive the PMF of the load, we will first focus on the characterization of the total length of the lines of the PLP inside the typical cell, which will be referred to as the \textit{total chord length} and denoted by $W$.
The distribution of the length of the chord segment of the line passing through the typical PV cell has already been provided in \cite[Section 4]{muche}. Therefore, we state this result in the following lemma without a proof. 

\begin{lemma}\label{lem:pdfc}
	The probability density function (PDF) of the chord length of the line intersecting the typical cell $V_o$ is $ f_C (c) $ 
	\begin{align}
	 &= \frac{\pi}{2} \lambda_b^{\frac{3}{2}} \int_0^\infty \mspace{-12mu} \int_0^\pi \mspace{-6mu} \tau \bigg[ \lambda_b \left(\frac{\partial K_{\tau, \alpha} }{\partial c } \right)^2 \scalebox{2}[1]{-}  \frac{\partial^2 K_{\tau, \alpha} }{\partial c^2 } \bigg] e^{- \lambda_b K_{\tau, \alpha} } \nrmd \alpha \nrmd \tau,\\
	\notag &{where} \quad	K_{\tau, \alpha} = 2 \pi \tau^2  - 2 \pi \tau c \cos \alpha - \tau^2 \left(\alpha - 0.5 \sin 2\alpha \right)\\
	& \quad + \pi c^2 - (\tau^2 -2 \tau c \cos \alpha + c^2) \left(\phi -  0.5 \sin 2\phi \right),\\
	 &{and} \hspace{1em} \phi = \arccos \left[ \frac{2 c^2 - 2 \tau c \cos \alpha }{2 c \sqrt{\tau^2 - 2 \tau c \cos \alpha + c^2 }}\right].
	\end{align}
\end{lemma}
Note that the number of lines that intersect the typical cell is random. Using the properties of the PLP and the PDF of the chord length given in the above lemma, we will now characterize the total chord length distribution in the typical cell in the following lemma.

\begin{lemma}\label{lem:lapl_chordlength}
	The Laplace transform of the total chord length distribution is 
	\begin{align}
	\notag &\calL_W(s) = \frac{1}{1163} \lambda_b^{3.7903} \int_0^\infty \mspace{-12mu} u^{6.5806}\exp \bigg[ - \lambda_l u  \\
	&  \times  \int_0^\infty \mspace{-12mu} \left(1 - e^{-sc} \right)  f_C (c) \nrmd c -0.1165 (\sqrt{\lambda_b} u)^{2.3361}\bigg] \nrmd u,
	\end{align}
	where $f_C(c)$ is given in Lemma \ref{lem:pdfc}.
\end{lemma}
\begin{IEEEproof}
	First, we compute the Laplace transform of the distribution of the total chord length, denoted by $W$, conditioned on the number of lines $N_0$ that intersect the typical cell $V_o$. Thus, we have	
	\begin{align} 
	\notag &\calL_W(s | N_0) = \nbbE \left[ e^{-s \sum_{i=1}^{n_0}  C_i } \mid N_0 = n_0\right] \\
	 & \stackrel{(a)}{=} \prod_{i=1}^{n_0} \nbbE \left[ e^{-s C_i}  \mid N_0 = n_0\right] 
	 \stackrel{(b)}{=}  \left[ \int_0^\infty \mspace{-12mu} e^{-s c}f_C (c) \nrmd c \right]^{n_0},
	\end{align} 
	where $C_i$ denotes the chord length of the line $L_i$ that intersects the typical cell, and (a) and (b) follow from the independent and identical distribution of chord lengths of different lines intersecting the typical cell.
	
	In order to calculate the Laplace transform of the total chord length, we must now compute the expectation of the above expression w.r.t. $N_0$. From the properties of the PLP, it follows that the number of lines of a motion-invariant PLP with line density $\mu$ that intersect a convex body $K$ follows a Poisson distribution with mean $\frac{\mu}{\pi} S_K$, where $S_K$ denotes the perimeter of $K$ \cite[Section II]{vishnuJ2}, \cite[Example 8.2]{stoyan}. Note that the perimeter of the typical cell is also random and its empirical distribution is well-known in the literature \cite{tanemura}. Thus, conditioned on the perimeter $U$ of the typical cell $V_o$, the Laplace transform of the chord length distribution is obtained as
	\begin{align}
	\notag \calL_W&(s | U )   = \nbbE_{N_0} \left[ \left( \int_0^\infty \exp(-s c) f_C (c) \nrmd c \right)^{n_0} \biggm| U\right] \\
	\notag & = \sum_{n_0 = 0}^\infty \frac{ \exp\left( - {\lambda_l u}\right) (\lambda_l u)^{n_0}}{n_0!}\left( \int_0^\infty \mspace{-6mu} \exp(-s c) f_C (c) \nrmd c \right)^{n_0} \\	
	\label{eq:cond_LW} & \stackrel{(a)}{=} \exp \left[ - \lambda_l u  + \lambda_l u \int_0^\infty \exp(-s c) f_C (c) \nrmd c \right] ,
	\end{align}
	where (a) follows from the Taylor series formula for the exponential function. From \cite{tanemura}, the PDF of the perimeter of the typical cell is given by
	\begin{align}
	&f_U(u) = \frac{\sqrt{\lambda_b}}{4} g\left(2.33609, 2.97006, 7.58060, \frac{\sqrt{\lambda_b}u}{4} \right),\\
	&\text{where }g(a,b,c,x) = {a b^{c/a}}{(\Gamma\left(c/a\right))^{-1}} x^{c-1} e^{ - b x^a}.
	\end{align}
	Upon computing the expectation of the expression in \eqref{eq:cond_LW} w.r.t. $U$, we obtain the final expression. 
\end{IEEEproof}

\subsection{Load distribution}

Using the Laplace transform of the distribution of the total chord length, we will now derive the PMF of the load on the typical MBS in the following theorem.

\begin{thm}\label{thm:load_typ_exact}
	The PMF of the load on the typical MBS is
	\begin{align}\label{eq:pmf_load_exact}
	\P (M = m) = \frac{(-\lambda_v)^m}{m!} \left[ \frac{\partial^m }{\partial s^m} \calL_W(s) \right]_{s= \lambda_v},
	\end{align}
	where $\calL_W(s)$ is given in Lemma \ref{lem:lapl_chordlength}.
\end{thm}
\begin{IEEEproof}
	Recall that the distribution of points on each line of the PLCP $\Phi_v$ follows a PPP with density $\lambda_v$. Therefore, due to the independent spatial distribution of points on different lines, the total number of points of $\Phi_v$ that lie inside the typical cell $V_o$ follows a Poisson distribution with mean $\lambda_v W$. Thus, conditioned on $W$, the PMF of the load is given by
	\begin{align}
	\P (M = m \mid W) = \frac{\exp(- \lambda_v w) (\lambda_v w )^m }{m!}.
	\end{align}
	By taking the expectation of the above expression w.r.t. $W$, we obtain the PMF of the load as
	\begin{align}
	\P (M =m ) = \int_0^\infty \frac{\exp(- \lambda_v w) (\lambda_v w )^m }{m!} f_W (w) \nrmd w.
	\end{align}
	Note that it is not straightforward to obtain the PDF of $W$. However, the above expression can be expressed in terms of the Laplace transform of $W$ as $\P (M =m )$
		\begin{align}
		\notag &  = \int_0^\infty \frac{(-\lambda_v)^m}{m!}  \left[ \frac{\partial^m e^{-sw}}{\partial s^m}\right]_{s= \lambda_v}  \mspace{-24mu} f_W (w) \nrmd w \stackrel{(a)}{=} \frac{(\scalebox{2}[1]{-}\lambda_v)^m}{m!} \\
		\notag  & \times \left[ \frac{\partial^m }{\partial s^m} \int_0^\infty \mspace{-12mu} e^{-s w} f_W (w) \nrmd w \right]_{s= \lambda_v} \mspace{-24mu} \stackrel{(b)}{=} \frac{(\scalebox{2}[1]{-}\lambda_v)^m}{m!} \left[ \frac{\partial^m \calL_W(s)}{\partial s^m}  \right]_{s= \lambda_v}\mspace{-12mu},
		\end{align}
	where (a) follows from Leibniz's rule for derivative of integral, and (b) follows from the definition of Laplace transform. {Note that the PMF for the case of $m=0$ yields the probability that there are no users in the typical cell and is simply obtained using $\calL_W(s)$ (i.e., without computing its derivative).}
\end{IEEEproof}

\subsection{Approximation of the PMF}
While the derivation of the exact PMF of the load concretely captures the subtle dependence of load on other sources of randomness, such as the perimeter of the typical cell, the resulting expression is not very conducive for numerical evaluation. Consequently, it is much more difficult to evaluate other metrics that utilize this result. Motivated by this, we propose an accurate approximation of the PMF of the load using the following assumption.

\begin{assumption}\label{assum:disc}
	We assume that the distribution of number of users in the typical cell centered at the origin is the same as that of the number of users inside a disc $B(o, R_t)$ centered at the origin with radius $R_t$ such that the area of the disc is equal to that of the typical cell, i.e. $\pi R_t^2 = | V_o|$.
\end{assumption}


{This assumption is inspired by the asymptotic result that the large PV cells are circular \cite[Theorem 4]{praful}. As will be shown in Section \ref{sec:results}, the PMF of the load computed using this approximation is remarkably accurate in the non-asymptotic regime as well.} Using this assumption, we derive the approximate PMF of the load in the following theorem.
\begin{thm}\label{thm:load_typ_approx}
	Under Assumption \ref{assum:disc}, the PMF of the load on the typical MBS is given by
	\begin{multline}
	\P ( M =m ) \approx  \frac{(-\lambda_v)^m}{m!}\Bigg[\frac{\partial^m}{\partial s ^m} \int_0^\infty  \exp\bigg[ - 2 \pi \lambda_l r_t\\
	\quad +2 \pi \lambda_l \int_0^{r_t}  e^{-2s \sqrt{r_t^2 - \rho^2} }  \nrmd \rho \bigg] f_{R_t}(r_t) \nrmd r_t \Bigg]_{s = \lambda_v} .
	\end{multline}
\end{thm}
\begin{IEEEproof}
	As we have shown earlier, the first step in the computation of the PMF of the load on the typical MBS is the characterization of the total chord length. For a line $L$ at a perpendicular distance $\rho < R_t$ from the origin, the length of the chord that intersects $ B(o,R_t) $ is $2 \sqrt{R_t^2 - \rho^2}$. From the properties of PLP, we also know that the number of lines of $\Phi_l$ that intersect the disc $B(o,R_t)$ follows a Poisson distribution with mean $ 2 \pi \lambda_l R_t $. Thus, conditioned on the radius of the disc $R_t$, the Laplace transform of the total chord length $W$ in $B(o, R_t)$ can be computed as
	\begin{align}
	\notag &\calL_W(s | R)  \approx \nbbE \left[ e^{ -s \sum\limits_{L \in \Phi_l}  \nu_1 (B(o,r_t) \cap L) }\right] = \nbbE \bigg[ \mspace{-12mu} \prod_{\substack{(\rho, \theta) \in \Phi_\calC \\ \rho \leq r_t}} \mspace{-20mu} e^{ - 2 s \sqrt{r_t^2 - \rho^2}}\bigg]\\
	\notag &  \stackrel{(a)}{=} \exp\left[ - 2 \pi \lambda_l \int_0^{r_t} 1 - \exp(-2s \sqrt{r_t^2 - \rho^2})  \nrmd \rho \right],
	\end{align}
	where (a) follows from the PGFL of the PPP $\Phi_\calC$. From \cite{tanemura}, the PDF of the area of the typical cell $Z$ is given by
	\begin{align}
	f_Z(z) = \lambda_b g(1.07950, 3.03226, 3.31122, \lambda_b z ). 
	\end{align}
	Since $R_t = \sqrt{|V_o|/ \pi}$, the PDF of the radius of the disc is $f_{R_t}(r_t) = 2 \pi r_t f_Z(\pi r_t^2)$.
	Thus, the Laplace transform of the total chord length distribution is 
	\begin{align}\label{eq:lapW_approx}
	  &\calL_W(s) \approx \int_0^\infty \mspace{-12mu}  \exp\bigg[ \scalebox{2}[1]{-} 2 \pi \lambda_l \int_0^{r_t}  \mspace{-6mu} 1\scalebox{2}[1]{-} e^{\scalebox{2}[1]{-}2s \sqrt{r_t^2 \scalebox{2}[1]{-} \rho^2}}  \nrmd \rho \bigg] f_{R_t}(r_t) \nrmd r_t.
	\end{align}
	From Theorem \ref{thm:load_typ_exact}, we know that the PMF of the load can be expressed in terms of the derivative of $\calL_W(s)$. Hence, upon substituting \eqref{eq:lapW_approx} in \eqref{eq:pmf_load_exact}, we obtain the approximate PMF of the load on the typical MBS.
\end{IEEEproof}

%

\section{Load on the tagged MBS}
In this section, we will derive the PMF of the load on the tagged MBS that serves the typical vehicular user. Without loss of generality, we translate the origin to the location of the typical vehicular user. Thus, under the Palm distribution of the PLCP, the resulting point process of vehicular users $\Phi_{v_0}$ is the superposition of $\Phi_v$, a 1D PPP $\Psi_{L_0}$ with density $\lambda_v$ along the line $L_0$ that passes through the origin, and an atom at the origin, i.e., $\Phi_{v_0} = \Phi_v \cup \Psi_{L_0} \cup \{o\}$. The line $L_0$ will henceforth be referred to as the typical line. The Voronoi cell of the tagged MBS that contains the origin is called the \textit{zero cell} and it is mathematically expressed as $ V(o) =$
\begin{align*}
 \{ \nrmy \in \nbbR^2 : \| \nrmy - \nrmx \| \leq \| \nrmy - \nrmx'\|,\  \|\nrmx\| < \| \nrmx'\|, \forall \nrmx' \in \Phi_b  \}.
\end{align*}

{Owing to the Palm distribution of the PLCP and the statistically larger area of the zero cell, its total chord length distribution is different from that of the typical cell.} So, we first derive the Laplace transform of the distribution of the total chord length in the following lemma by using an assumption similar to that of Assumption \ref{assum:disc}.

\begin{lemma}\label{lem:lapl_chordlength_zerocell}
	The Laplace transform of the distribution of the total chord length in the zero cell is
	\begin{align}
	\notag&\calL_W(s) \approx \int_0^\infty \int_0^\infty \exp \bigg[ -s c_0 - 2 \pi \lambda_l r_z + 2 \pi \lambda_l \\
	&\times  \int_0^{r_z} e^{-2s \sqrt{r_z^2 - \rho^2}}  \nrmd \rho \bigg] f_{R_z}(r_z) f_{C_0}(c_0) \nrmd c_0 \nrmd r_z,
	\end{align}
	where $f_{R_z}(r_z) = 2 \pi^2 \lambda_b r_z^3 f_Z(r_z)$, and $f_{C_0} = \frac{4 \sqrt{\lambda_b}}{\pi} c_0 f_C(c_0)$.
\end{lemma}
\begin{IEEEproof}
	We know that the typical line $L_0$ intersects the zero cell and we denote the corresponding chord length by $C_0$. In addition to this, there exists a random number of lines of $\Phi_l$ that intersect the zero cell. Thus, we write the total chord length of the zero cell as $W = C_0 + C_1$, where $C_1$ denotes the sum of chord lengths of all lines that intersect $V(o)$ excluding the typical line. We will compute the Laplace transforms of the distribution of $C_0$ and $C_1$ separately.
	 
	The PDF of the typical chord length of a line in an arbitrarily chosen PV cell is given in Lemma \ref{lem:pdfc}. However, we are interested in the length of the chord segment that contains the origin, which is expected to be longer than the length of the typical chord segment. Thus, using length-biased sampling, the PDF of $C_0$ is given by
	\begin{align}
	f_{C_0}(c_0)  =  \frac{c_0 f_{C}(c_0)}{\nbbE[C]} = \frac{4 \sqrt{\lambda_b}}{\pi} c_0 f_C(c_0). 
	\end{align}
	Using this result, the Laplace transform of the distribution of $C_0$ can be computed as
	\begin{align}\label{eq:lapC0}
	\calL_{C_0}(s) = \int_0^\infty \exp(-s c_0) f_{C_0}(c_0) \nrmd c_0.
	\end{align}
	
	We will now focus on the computation of the Laplace transform of the distribution of $C_1$. First, we assume that the number of vehicular users in the zero cell, excluding the users on the typical line, is the same as the number of vehicular users in a disc $B(\nrmx, R_z)$ centered at the location of the serving MBS $\nrmx$ with radius $R_z$ such that $\pi R_z^2 = |V(o)| $. From area-biased sampling \cite{Dhi_letter}, the PDF of the area of the zero cell is 
	\begin{align}
	f_{Z'}(z') = \frac{z' f_Z(z')}{\nbbE[Z]} = \lambda_b z' f_Z(z'). 
	\end{align}
	
	Following the same procedure as in the proof of Theorem \ref{thm:load_typ_approx}, we obtain the approximate Laplace transform of the distribution of $C_1$ as $\calL_{C_1}(s) $
	\begin{align}\label{eq:lapC1}
	 \approx \mspace{-6mu} \int_0^\infty \mspace{-12mu} \exp\bigg[ - 2 \pi \lambda_l \int_0^{r_z} \mspace{-12mu}  1-e^{-2s \sqrt{r_z^2 - \rho^2} }  \nrmd \rho \bigg] f_{R_z}(r_z) \nrmd r_z,
	\end{align}
	where $f_{R_z}(r_z) = 2 \pi r_z f_{Z'}(\pi r_z^2)$.
	
	 As $C_0$ and $C_1$ are independent, the Laplace transform of the distribution of total chord length can be computed as 
	\begin{align}\label{eq:lapW_zerocell}
	\calL_W(s) = \calL_{C_0}(s)  \calL_{C_1}(s).
	\end{align}
	Substituting \eqref{eq:lapC0} and \eqref{eq:lapC1} in the above equation yields the expression for the Laplace transform of the total chord length distribution in the zero cell.
	\end{IEEEproof}

\begin{thm}\label{thm:load_tagged}
	The PMF of the load on the tagged MBS is 
	\begin{align}
		\notag &\P (M = m+1) = \frac{(-\lambda_v)^m}{m!} \int_0^\infty \mspace{-12mu} \int_0^\infty \Bigg[  \frac{\partial^m }{\partial s^m} \exp \bigg[-s c_0 \\
		\label{eq:pmf_zerocell} & \scalebox{2}[1]{-}2 \pi \lambda_l  \int_0^{r_z} \mspace{-12mu} 1\scalebox{2}[1]{-} e^{-2s \sqrt{r_z^2 - \rho^2}}  \nrmd \rho \bigg]  \Bigg]_{s= \lambda_v} \mspace{-28mu}f_{R_z}(r_z) f_{C_0}(c_0) \nrmd c_0 \nrmd r_z,
	\end{align}
	where $ m = 0,\ 1,\ 2, \dots$.
\end{thm}
\begin{IEEEproof}
	The proof follows along the same lines as that of Theorem \ref{thm:load_typ_exact}.
\end{IEEEproof}

\section{Results and Discussion}

In this section, we will verify the accuracy of our analytical results and also discuss the application of these results in the computation of rate coverage of the typical vehicular user. 
\begin{figure}
	\centering
	\includegraphics[width=0.4\textwidth]{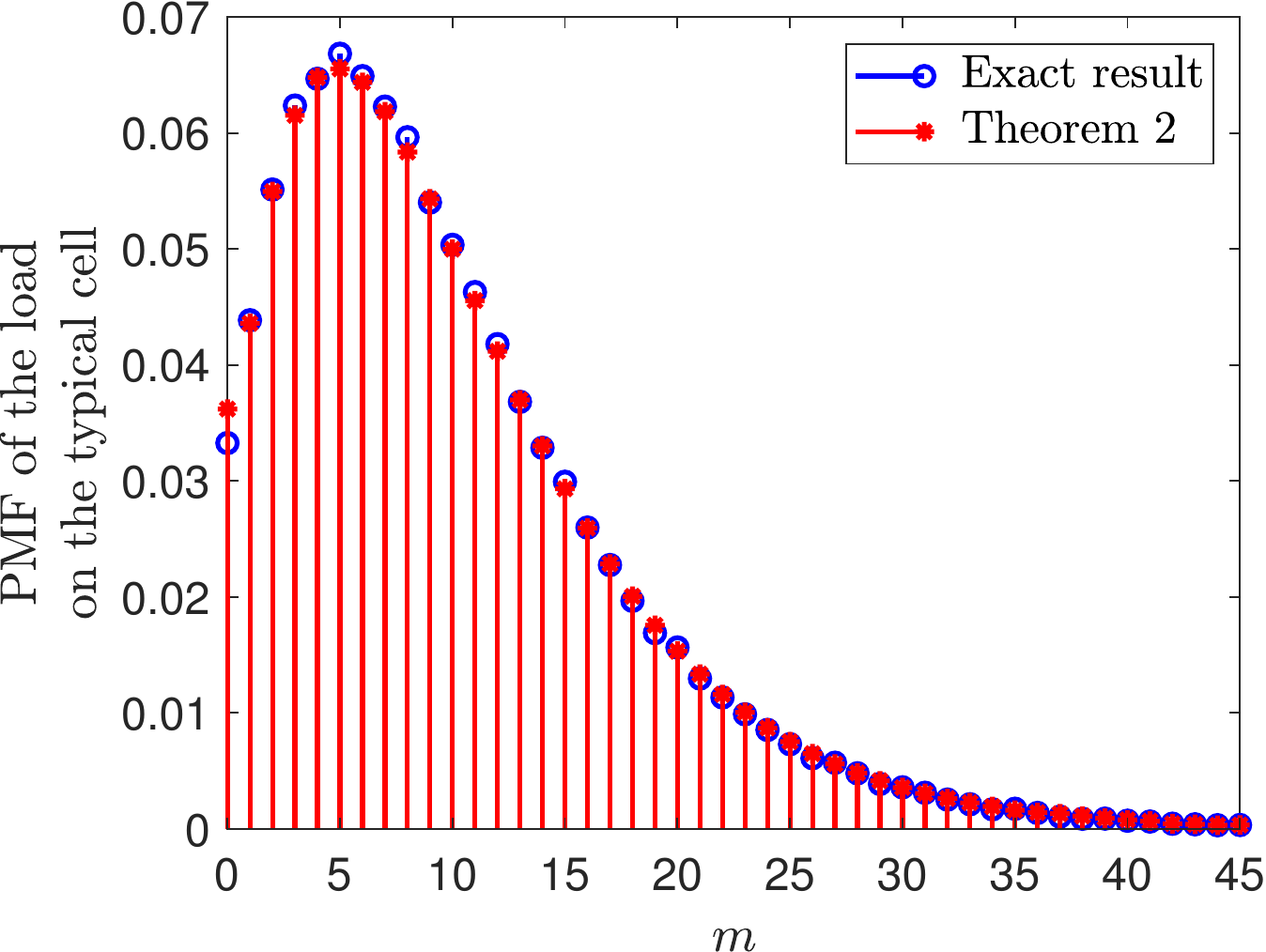}
	\caption{{PMF of the load on the typical MBS ($\mu_l = 5$ km$^{-1}$, $\lambda_v = 2$ nodes/km, and $\lambda_b = 1$ node/km$^2$).}}
	\label{fig:pmf_typ}
\end{figure}
\begin{figure}
	\centering
	\includegraphics[width=0.4\textwidth]{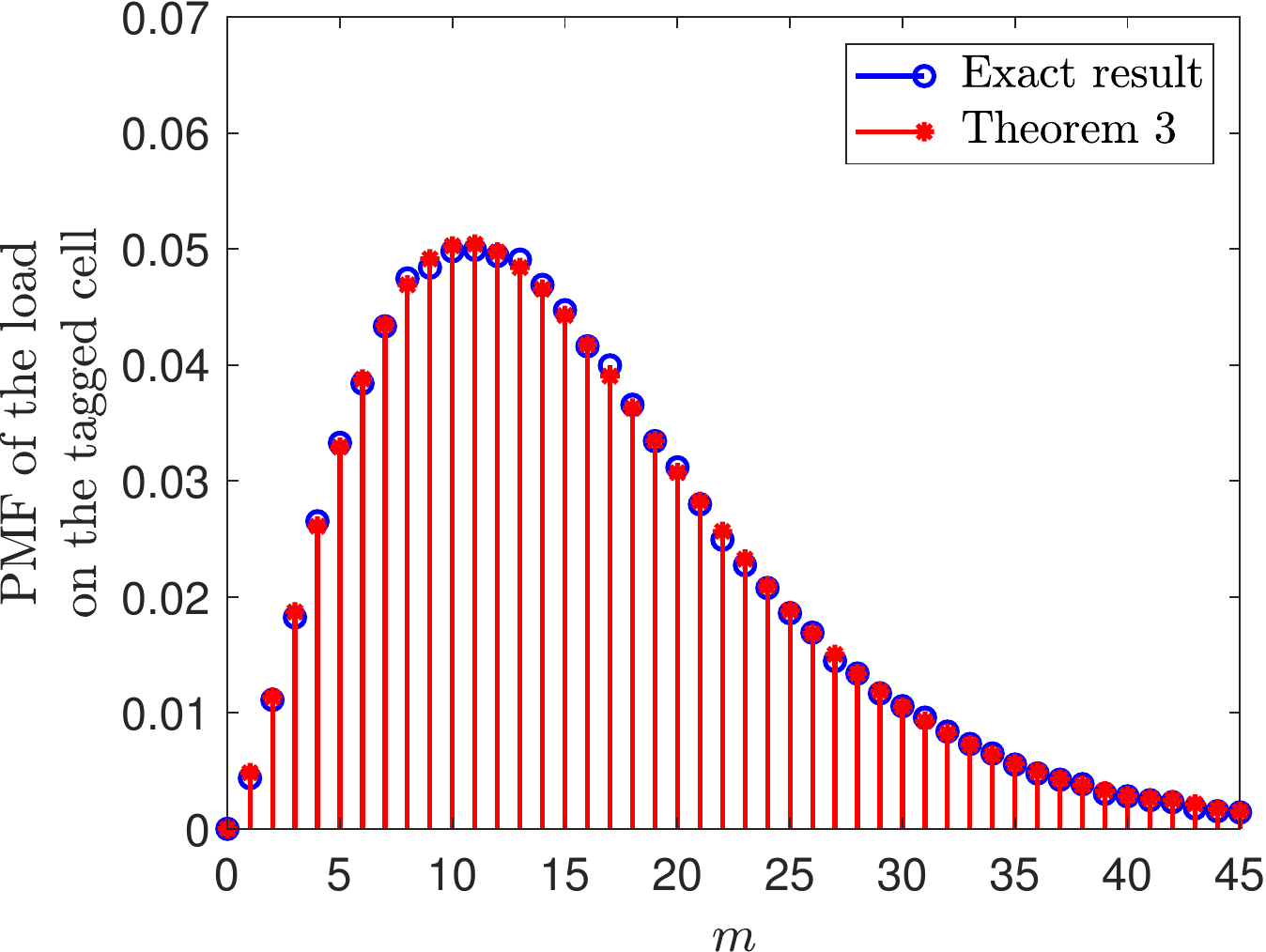}
	\caption{{PMF of the load on the tagged MBS ($\mu_l = 5$ km$^{-1}$, $\lambda_v = 2$ nodes/km, and $\lambda_b = 1$ node/km$^2$).}}
	\label{fig:pmf_zero}
\end{figure}

\subsection{Numerical results}\label{sec:results}
In Fig. \ref{fig:pmf_typ}, we plot the PMF of the load on the typical cell evaluated from expressions given in Theorem \ref{thm:load_typ_approx} along with the empirical PMF of the load obtained through Monte-Carlo simulations. It is quite evident that the results derived in Theorem \ref{thm:load_typ_approx} match closely with the empirical results from simulations thereby demonstrating the accuracy of our analytical results. Similarly, the PMF of the load on the tagged MBS evaluated using the expressions in Theorem \ref{thm:load_tagged} matches closely with the simulations as shown in Fig. \ref{fig:pmf_zero}.

\subsection{Rate coverage}
In this subsection, we will focus on the characterization of rate coverage of the typical receiver in the network using the load distribution of vehicular users. We consider a single tier vehicular network as described in Section \ref{sec:sysmod}. Without loss of generality, we translate the origin to the location of the typical receiver. Thus, the resulting point process of vehicular users is $\Phi_{v_0} \equiv \Phi_v \cup \Psi_{L_0} \cup \{o\}$. We assume that all the MBSs have the same transmit power $P_t$ and the typical receiver connects to the closest MBS in the network. In this setup, the SIR measured at the typical receiver is 
\begin{align}
\sir = \frac{P_t H_0 R^{-\alpha}}{\sum_{\nrmx \in \Phi_b} P_t H_\nrmx \| \nrmx \|^{-\alpha} },
\end{align}
where $R$ denotes the serving distance, and $H_0$ and $H_\nrmx$ denote the fading gains of the links from the typical receiver to the serving MBS and interfering MBSs at $\nrmx$, respectively, which are exponentially distributed with unit mean. Assuming that the bandwidth $B$ is equally shared by all the users served by the serving MBS, the achievable rate at the typical receiver is 
\begin{align}\label{eq:rate}
\calR = \frac{B}{M}\log_2(1 + \sir),
\end{align}
where $M$ denotes the load on the serving MBS. We present the rate coverage of the typical receiver, which is defined as the probability with which the rate at the typical receiver exceeds a desired threshold $T$, in the following theorem.

\begin{thm}\label{thm:rate_coverage}
	The rate coverage of the typical receiver is $ \mathsf{R}_{\rm c} $
	\begin{align*}
	 & \approx \sum_{m=1}^\infty P_m 2\pi\lambda_b \int_0^\infty \mspace{-12mu} r\exp \bigg[ \scalebox{2}[1]{-}\lambda_b \pi r^2   \scalebox{2}[1]{-}  \int_r^\infty \mspace{-6mu} \frac{2 \pi \lambda_b \gamma y}{\gamma  +  (y/r)^\alpha } \nrmd y \bigg] \nrmd r,
	\end{align*}
	where $\gamma = 2^{\frac{T m}{B}} -1$, and $P_m \equiv \P(M = m) $ is given by Theorem \ref{thm:load_tagged}.
\end{thm}
\begin{IEEEproof}
	Although $M$ and $\sir$ in \eqref{eq:rate} are negatively correlated in general, we consider them to independent in the interest of analytical tractability. This is a popular assumption in the literature and its accuracy has been verified in many settings, e.g., see \cite{Dhi_letter}. Thus, the rate coverage of the typical receiver can be computed as 
	\begin{align}
	\notag \mathsf{R}_{\rm c} &= \P ( \calR > T) = \P \left( \frac{B}{M}\log_2 (1 + \sir) > T\right)\\
	\label{eq:rc_step1} &=\sum_{m=1}^\infty \P(M = m) \P \left( \sir > 2^{\frac{T m}{B}} -1 \right).
	\end{align}
	For this setup, {the probability that $\sir$ at the typical receiver is greater than a desired threshold $\beta$, referred to as the \textit{coverage probability} of the typical receiver in an interference dominant setting,} is well-known in the literature \cite{DhiGanBacAnd} and is given by $\P(\sir>\beta)= $
	\begin{align}\label{eq:coverage}
	 &2\pi\lambda_b \int_0^\infty \mspace{-12mu} r\exp \bigg[ -\lambda_b \pi r^2  -  \int_r^\infty \frac{2 \pi \lambda_b \beta y}{ \beta +   r^{-\alpha}y^\alpha } \nrmd y \bigg] \nrmd r.
	\end{align}
	We also know the expression for the PMF of the load on the serving MBS from Theorem \ref{thm:load_tagged}. So, upon substituting \eqref{eq:coverage} and \eqref{eq:pmf_zerocell} in \eqref{eq:rc_step1}, we obtain the expression for rate coverage.
\end{IEEEproof}

\begin{figure}
	\centering
	\includegraphics[width=0.4\textwidth]{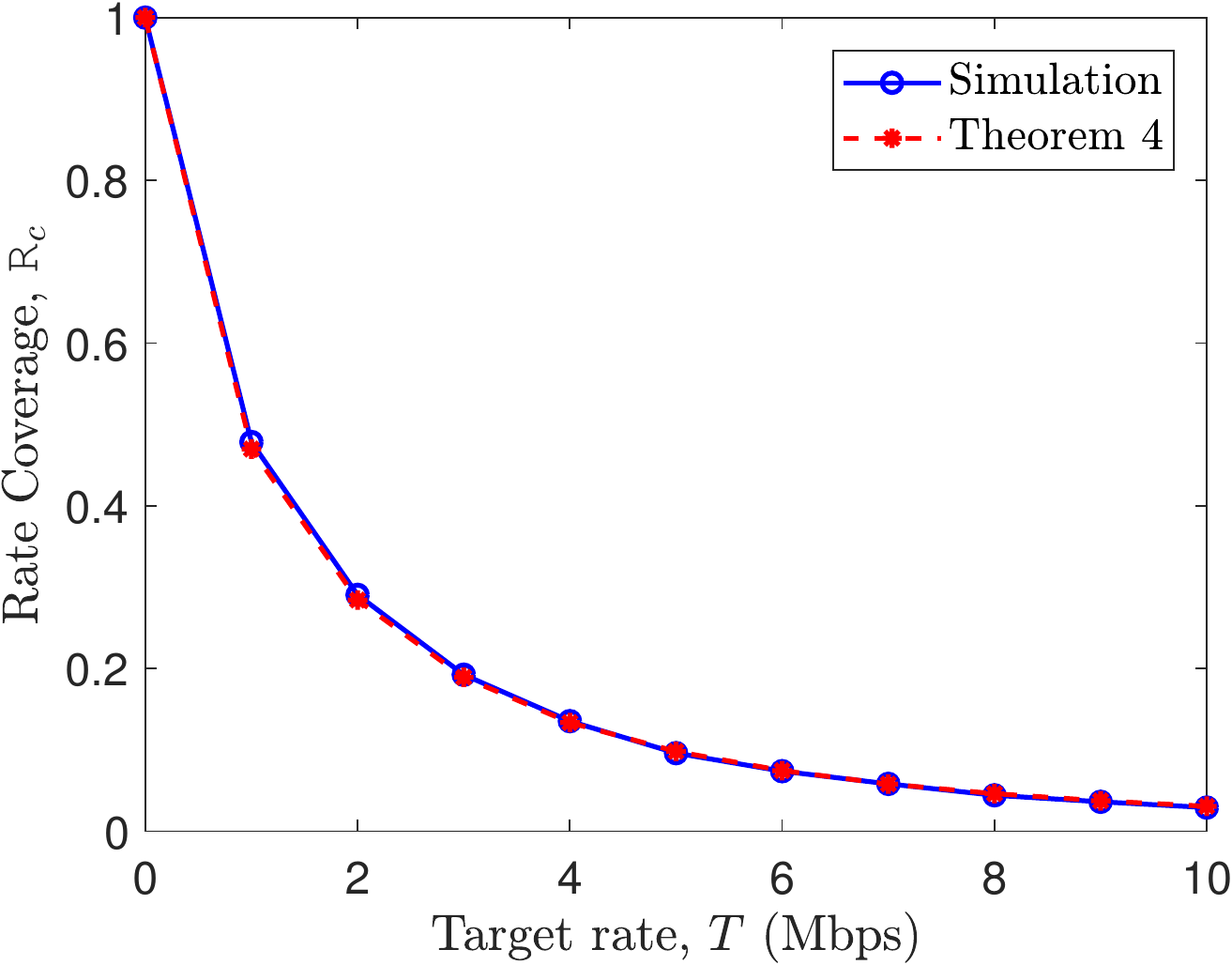}
	\caption{{Rate coverage of the typical receiver ($\mu_l = 5$ km$^{-1}$, $\lambda_v = 2$ nodes/km, $\alpha$ = 4, and $B$ = 10 MHz).}}
	\label{fig:rate_coverage}
\end{figure}

From Fig. \ref{fig:rate_coverage}, it can be noted that the rate coverage of the typical receiver evaluated using the analytical expression given in Theorem \ref{thm:rate_coverage} matches with the simulation results.

\section{Conclusion}
In this letter, we have derived the PMF of the load on cellular MBSs due to vehicular users modeled by PLCP in a single-tier network. Specifically, we characterized the total chord length distribution in a typical PV cell, which is an important result in its own right, and computed the PMF of the load on the typical MBS. As the evaluation of this expression is quite challenging, we proposed an accurate approximation for the PMF of the load on the typical MBS. Following a similar approach, we then derived the PMF of the load on the tagged MBS. Using this result, we also computed the rate coverage of the typical vehicular user in the network.

\bibliographystyle{IEEEtran}
\bibliography{Chetlur_WCL2020-0849_arxiv.bbl}

\end{document}